\documentstyle[multicol,aps,amssymb,prl,epsf]{revtex}
\begin{document}
\draft
\title{Hydrodynamics of topological defects in nematic liquid crystals}
\author{G\'eza T\'oth$^1$, Colin Denniston$^2$, and J. M. Yeomans$^1$}
\address{$^1$ Dept. of Physics, Theoretical Physics,
University of Oxford, 1 Keble Road, Oxford OX1 3NP}
\address{$^2$ Dept. of Physics and Astronomy, 
The Johns Hopkins University, Baltimore, MD 21218.}
\date{\today}
\maketitle

\begin{abstract} 
We show that back-flow, the coupling between the order parameter
and the velocity fields, has a significant effect on the motion of defects in
nematic liquid crystals. In particular the defect speed can
depend strongly on the topological strength in two dimensions and on the 
sense of rotation of the director about the core in three dimensions. 
\end{abstract}

\pacs{61.30.Jf,83.80.Xz,61.30.-v}

\begin{multicols}{2}

Topological defects arise in all areas of physics from cosmic strings 
\cite{T89} to vortices in superfluid helium \cite{O50}. Although the
physical systems are very different, many aspects of the observed
phenomena match even quantitatively, making it possible to test
cosmological predictions experimentally in condensed matter systems
\cite{T89}. Defects are classified according to their topological
strength, and in many cases, the symmetries of the field equations
lead to dynamics where defects of opposite topological strength
can be mapped into each other.  In this Letter we show an example, of
topological line defects in liquid crystals \cite{GP93}, where this
symmetry is broken, due to the coupling to an additional field, 
the flow-field, or to elastic constants that are not equal. 

In liquid crystals the topological defects are moving within a liquid 
and therefore
one must expect hydrodynamics to play an important role in their
dynamics. In particular, as the defects move, the coupling between the
changing director field and the velocity field (so-called back-flow) 
may play a significant part in the motion.
Experimental evidence \cite{AT00} shows that this is indeed the case when a
nucleated domain where the director field is horizontal
grows in a twist or vertical environment due to the
influence of the surfaces or an external electric field.  The speed of
the domain boundary is found to depend strongly on the local defect
configuration. 

Previous investigations \cite{JM96,PR92,P88,D96} of defect dynamics have 
either ignored the flow field or taken account of its effect
phenomenologically.  Here we aim to generalize this work by treating
the full hydrodynamic equations of motion for a nematic liquid
crystal. We consider the annihilation of a pair of defects of
strength $s=\pm1/2$. We find that back-flow can change the speed of 
defects by up to $\sim 100\%$. 
Defects of different strength couple to the flow field in
different ways. This leads to a dependence of speed on strength
which can occur either as a result of the back-flow or if the elastic
energy is treated beyond a one-elastic constant approximation. 

The hydrodynamics of liquid crystals is often well described by the
Eriksen-Leslie-Parodi equations of motion, which are written in terms of the
director field. However these are restricted to an uniaxial order parameter
of constant magnitude. Thus they are inadequate to explore the hydrodynamics
of topological defects where the magnitude of the order parameter has a steep
gradient and becomes biaxial within the core region \cite{SS87}.
Here we consider the more general Beris-Edwards \cite{BE94} formulation of
nematohydrodynamics, where the equations of motion
are written in terms of a tensor order parameter $\bf Q$.

The equilibrium properties of the liquid crystal are described by a
 Landau-de Gennes free energy density \cite{GP93}. This
 comprises a bulk term 
\begin{equation}
f_{b}=\frac {A}{2} (1 - \frac {\gamma} {3})
          Q_{\alpha \beta}^2 - 
          \frac {A \gamma}{3} 
          Q_{\alpha \beta}Q_{\beta \gamma}
          Q_{\gamma \alpha} + 
          \frac {A \gamma}{4} (Q_{\alpha \beta}^2)^2,
\label{eqBulkFree}
\end{equation}
which describes a first order transition from the isotropic to the
nematic phase at $\gamma=2.7$, together with an elastic contribution
\begin{eqnarray}
f_{d} &=&\frac{L_1}{2} (\partial_\alpha 
               Q_{\beta \gamma})^2+
               \frac{L_2}{2} (\partial_\alpha 
               Q_{\alpha \gamma})
               (\partial_\beta 
               Q_{\beta \gamma}) \nonumber \\
&+&\frac{L_3}{2} Q_{\alpha \beta}
              (\partial_\alpha Q_{\gamma \epsilon})
               (\partial_\beta Q_{\gamma \epsilon}),
\label{fFrank}
\end{eqnarray}
where the $L$'s are material specific elastic constants. 
The Frank expression for the elastic energy,
written in terms of the derivatives of the director, can be simply mapped to
(\ref{fFrank})\cite{BE94}. $A$ controls the relative magnitude of $f_b$ and
$f_d$. The Greek indices label the Cartesian components of $\bf Q$,
with the usual sum over repeated indices.

The dynamics of the order parameter is described by the equation
\begin{equation}
(\partial_t+{\vec u}\cdot{\bf \nabla})
           {\bf Q}-{\bf S}({\bf W},{\bf
  Q})= \Gamma {\bf H},
\label{Qevolution}
\end{equation}
where ${\vec u}$ is the bulk fluid velocity and ${\Gamma}$ is a collective
rotational diffusion constant.  The term on the right-hand side of equation
(\ref{Qevolution})  describes the relaxation of the order parameter towards
the minimum of the free energy ${\cal F}$
\begin{eqnarray}
{\bf H}&=& -{\delta {\cal F} \over \delta {\bf Q} }+({\bf
    I}/3) {\mbox Tr}\left\{ \delta {\cal F} \over \delta {\bf Q} \right\}.
\label{H(Q)}
\end{eqnarray}  
The term on the left-hand side is
\begin{eqnarray}
{\bf S}({\bf W},{\bf Q})&
=&(\xi {\bf D}+{\bf \Omega})({\bf Q}+{\bf I}/3)+({\bf Q}+
{\bf I}/3)(\xi {\bf D}-{\bf \Omega}) \nonumber \\
&&-2 \xi ({\bf Q}+{\bf I}/3){\mbox{Tr}}({\bf Q}{\bf W}),
\end{eqnarray}
where  ${\bf D}=({\bf W}+{\bf W}^T)/2$ and
${\bf \Omega}=({\bf W}-{\bf W}^T)/2$
are the symmetric part and the anti-symmetric part respectively of the
velocity gradient tensor $W_{\alpha\beta}=\partial_\beta u_\alpha$.
$\xi$ is related to the aspect ratio of the molecules.  

The velocity field ${\vec u}$ obeys
the continuity equation and a Navier-Stokes equation with a stress
tensor generalized to describe the flow of nematic liquid crystals 
\begin{eqnarray}
\sigma_{\alpha\beta} &=&-\rho T \delta_{\alpha \beta}
-(\xi - 1)H_{\alpha\gamma}(Q_{\gamma\beta}+{1\over
  3}\delta_{\gamma\beta})\nonumber\\
&-&(\xi+1) (Q_{\alpha\gamma}+{1\over
  3}\delta_{\alpha\gamma})H_{\gamma\beta}\nonumber\\
&+& 2\xi (Q_{\alpha\beta}+{1\over 3}
\delta_{\alpha\beta})Q_{\gamma\epsilon}
H_{\gamma\epsilon}
- \partial_\beta Q_{\gamma\nu} {\delta
{\cal F}\over \delta\partial_\alpha Q_{\gamma\nu}},
\label{BEstress}
\end{eqnarray}
where $\rho$ and $T$ are the density and temperature.
Notice that the stress (\ref{BEstress}) depends on the molecular field
${\bf H}$ and on ${\bf Q}$. This is the origin of back-flow.
Details of the equations of motion can be found in reference \cite{BE94}.
The equations (\ref{eqBulkFree})--(\ref{BEstress}) were 
solved numerically using a lattice Boltzmann algorithm 
described in \cite{DO00}. 

\begin{figure}
\narrowtext
\centerline{\epsfxsize=3.0in
\epsfbox{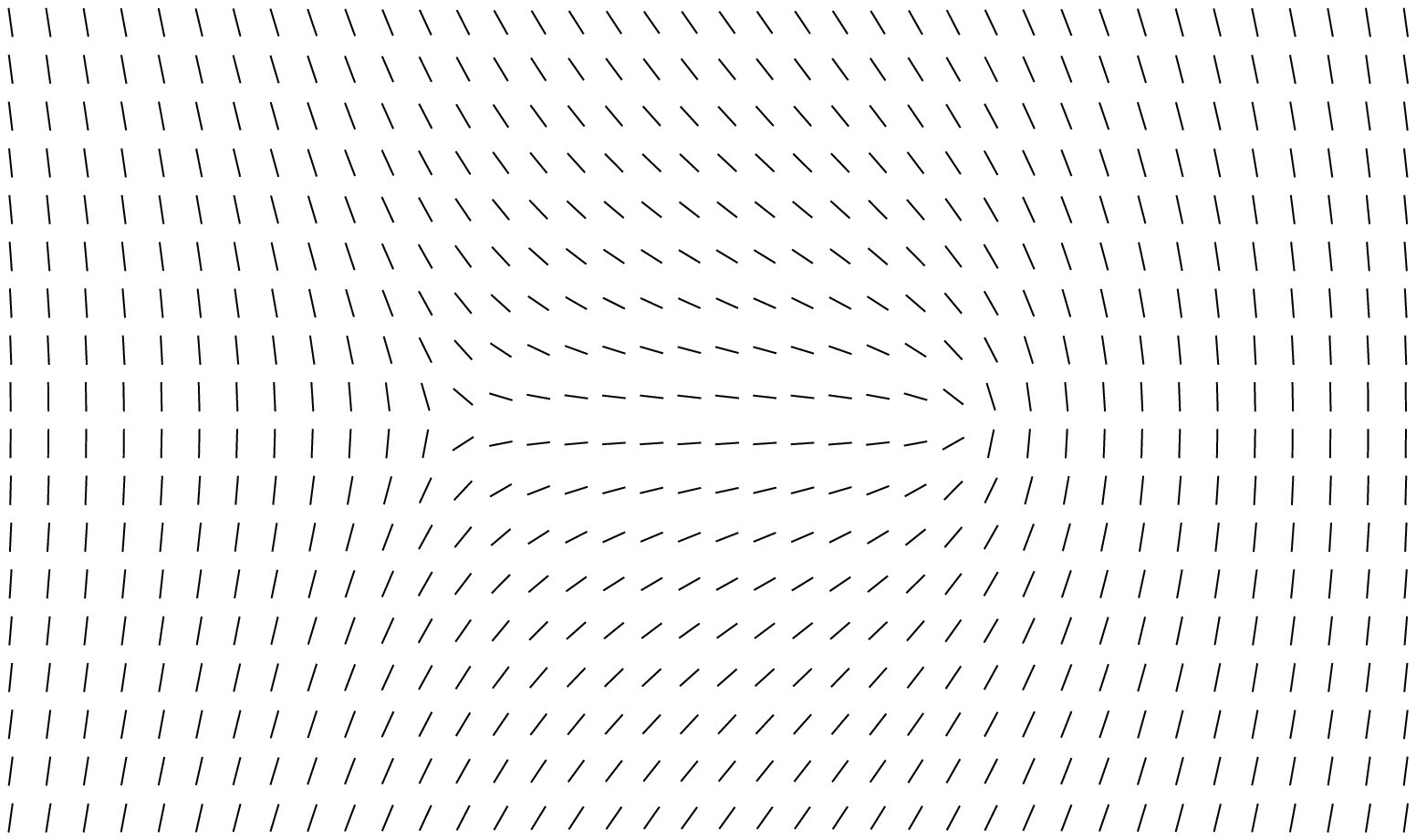}}
\centerline{(a)}
\centerline{\epsfxsize=3.0in
\epsfbox{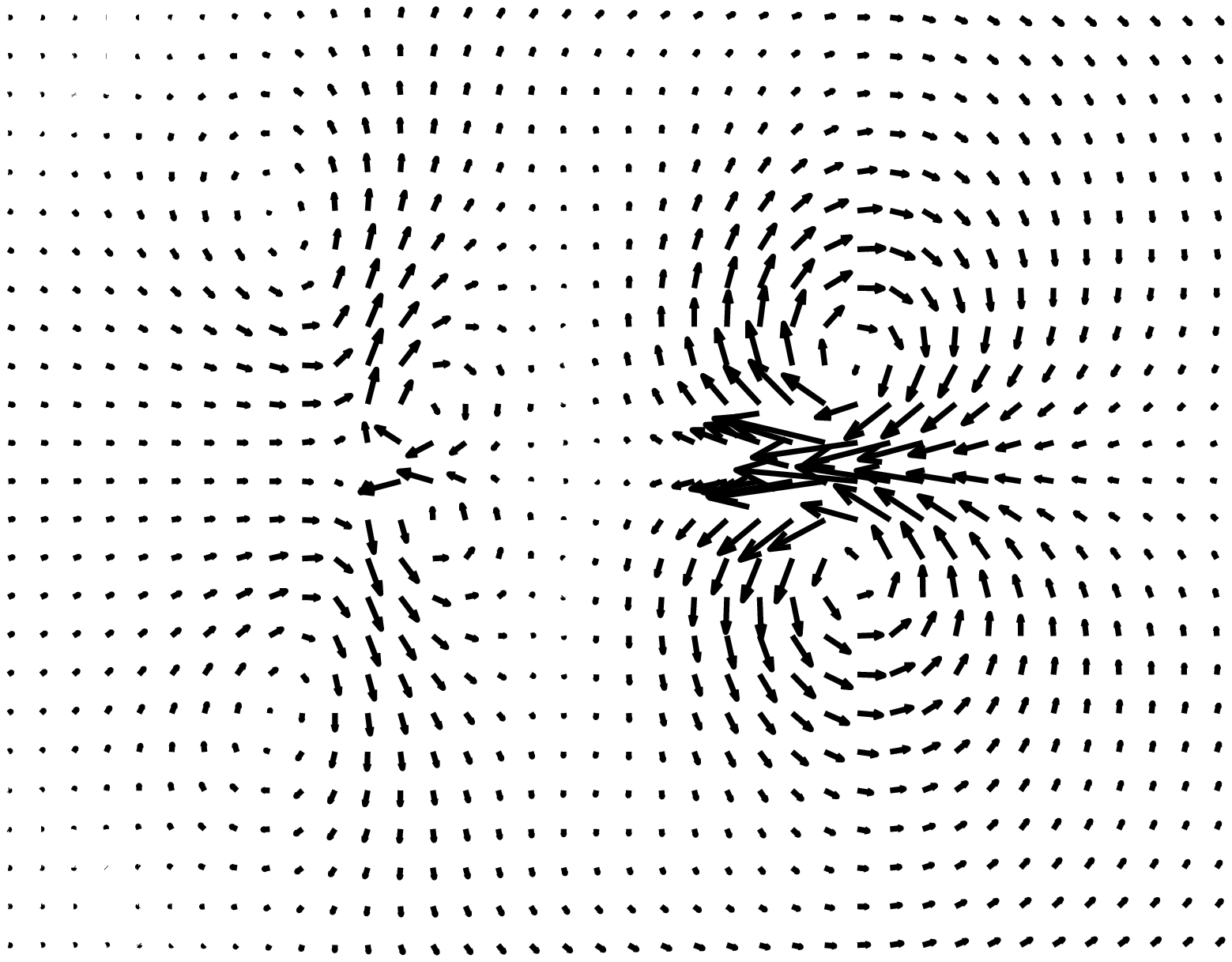}}
\centerline{(b)}
\caption{(a) The director field of two annihilating 
topological defects with strength $s= \pm 1/2$.  
(b) Velocity field of the two defects.} 
\label{Twodefects}
\end{figure}

Consider a pair of defects of topological strength $s=\pm 1/2$ situated
a distance $D$ apart in a nematic liquid crystal, as shown in 
Fig.~\ref{Twodefects}(a).  We consider a two-dimensional cross section of 
the two line defects, assuming that the order parameter does not change 
in the perpendicular direction (although the director may point out of this
plane).  The two defects are topologically distinct only in two dimensions, 
but even in three dimensions they are separated by an energy barrier
for the typical elastic constants we study here. 

A phenomenological equation of motion can be written down by assuming
that the attractive force between the two defects\cite {GP93} is
counterbalanced by a friction force\cite {P88}
\begin{equation}
D \frac{dD} {dt} = \mu_0 ln^{-1}(D/R_c),
\label{defectdyneq}
\end{equation}
where $D$ is the defect separation, $R_c$ is the defect core size
and $\mu_0$ is a constant.

In Ref. \cite{D96} the director field and the trajectory of the defects
were obtained analytically and the defect velocity was determined
as a function of parameters of the medium. (A review of the earlier
development of the theory of defect dynamics is also given in
\cite{D96}.)  However this and, as far as we are aware, all other analyses
of defect dynamics have ignored back-flow. 
This means that the approach to equilibrium is relaxational,
determined entirely by the derivative of the free energy with respect
to the order parameter, with the flow playing no role.
A further simplification in
previous work is that the Frank elastic constants were assumed to be
equal.

We can examine relaxational dynamics using a Ginzburg-Landau
equation for the director field \cite{B94}, i.e. Eqn.\
(\ref{Qevolution}) with the velocity set to zero.   
The Ginzburg-Landau equation with a single elastic constant is invariant
under a local coordinate transformation mirroring the director on the $x$
axis (where we define $x$ as the axis connecting the two defects cores).
This corresponds to the transformation
\begin{equation}
Q_{xy}\rightarrow -Q_{xy},\;\;\;\;\;\;\;\;\;\;\;\;
Q_{yx} \rightarrow -Q_{yx}.
\label{mapping}
\end{equation}
The order parameter fields of the two defects with topological 
charges $s=\pm 1/2$ transform into each
other. Thus approaches based on a simple Ginzburg-Landau equation predict
that when the defects move they follow symmetric dynamical
trajectories.

\begin{figure} 
\narrowtext
\centerline{\epsfxsize=3.0in\epsfbox{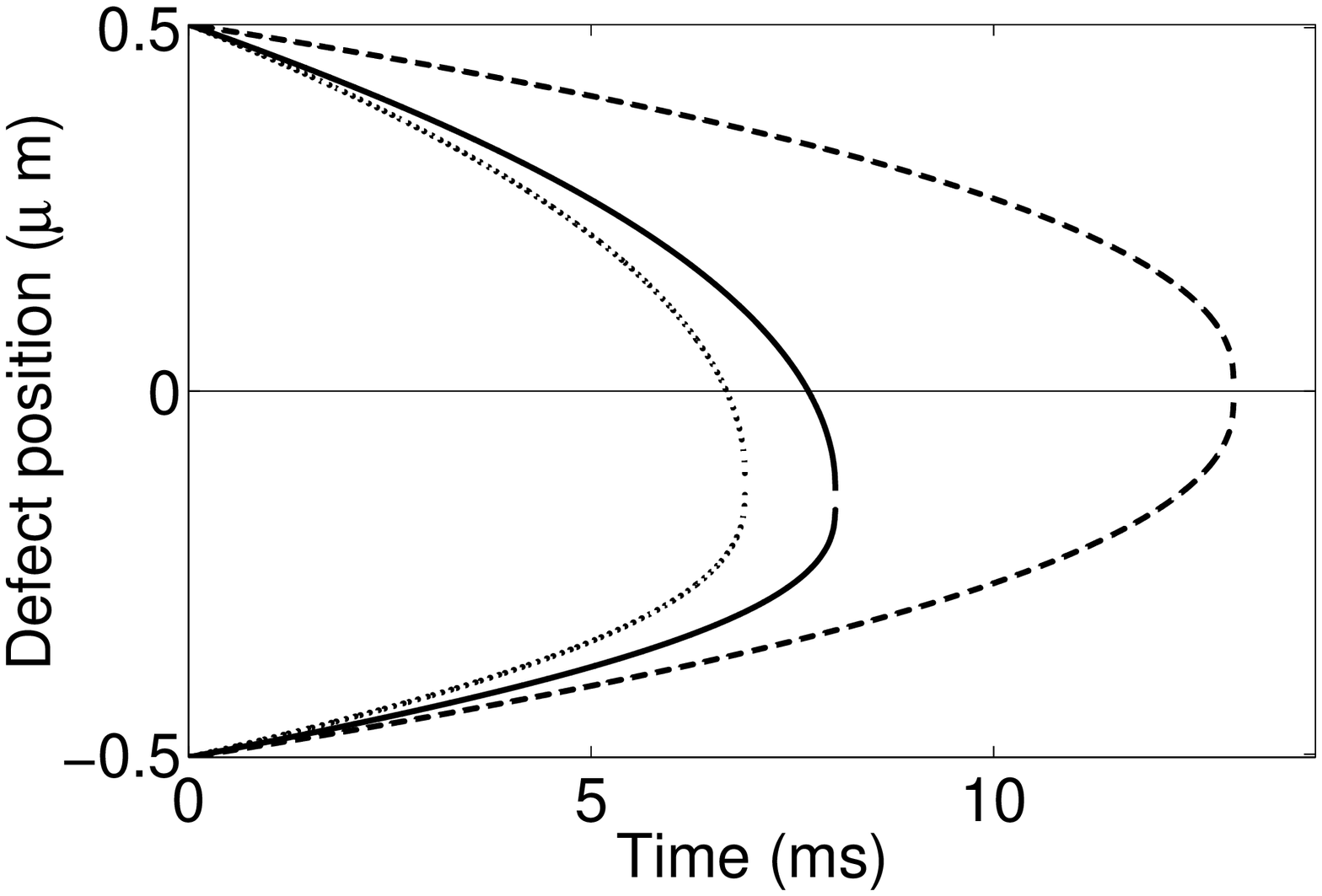}}
\centerline{\quad\epsfxsize=3.0in\epsfbox{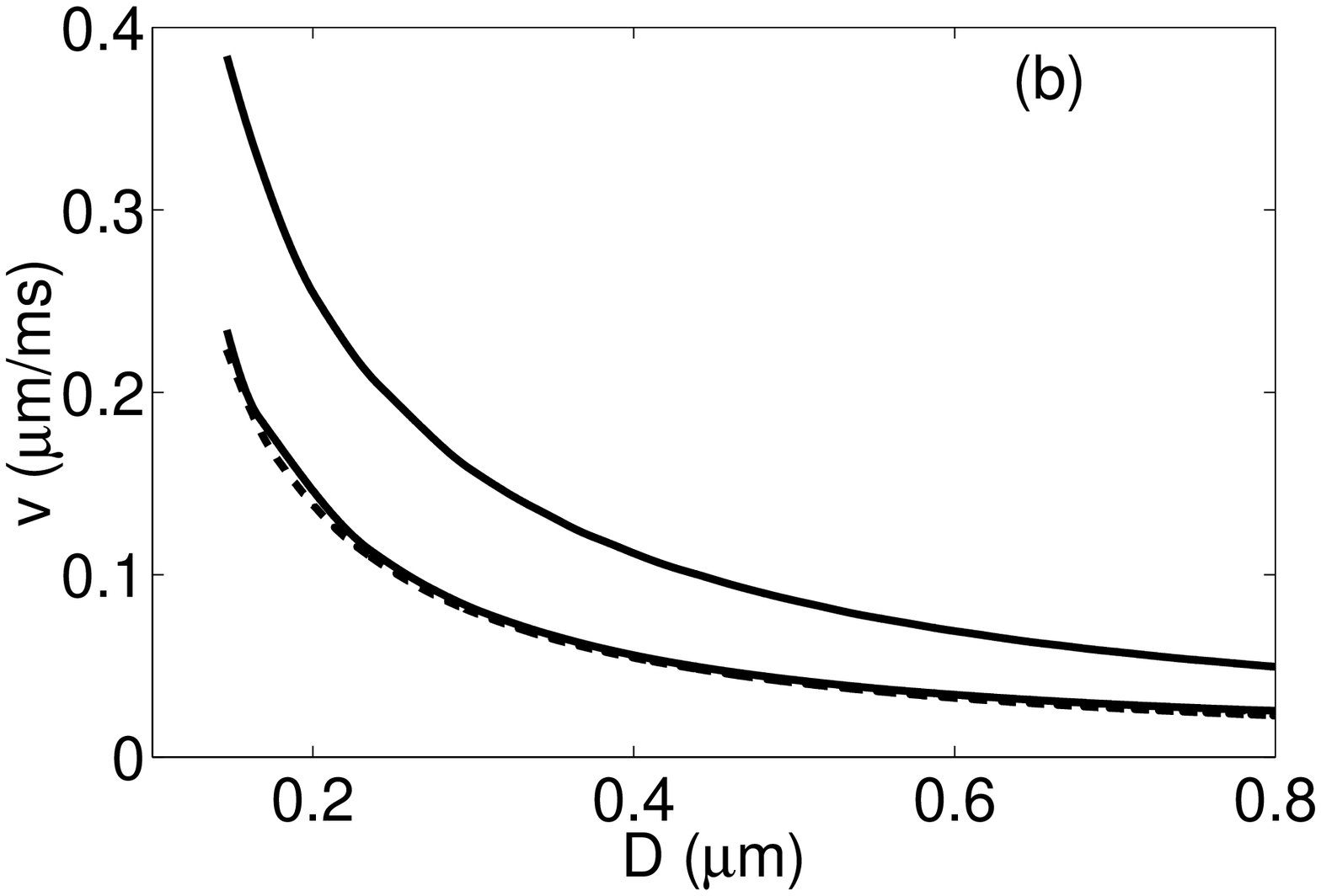}}
\caption{(a) Positions of the defects as a function of time.  Upper (lower)
  curves are for $s=+1/2$$(-1/2)$.   The different lines 
  correspond to a Ginzburg-Landau 
  model(dashed), the hydrodynamic equations of motion with 
  $\Gamma=6.25$ Pa$^{-1}$s$^{-1}$ (solid) and with 
  $\Gamma=7.76$ Pa$^{-1}$s$^{-1}$(dotted).  (b) Defect speed as a function 
  of separation.  Upper and lower solid curves: the $s=+1/2$ and
  $s=-1/2$ defect trajectories with hydrodynamics and $\Gamma=6.25$;
  Dashed curve: Ginzburg-Landau model. The $+1/2$ defect is
  considerably ($100\%$) accelerated for $D > 0.25 \mu$m compared to the
  results of the Ginzburg-Landau model. The speed of the $-1/2$ defect
  is only slightly affected by the back-flow.}
\label{Trajectories} 
\end{figure}

 Fig.\ \ref{Trajectories}(a) shows the position of two annihilating topological
defects, with topological charge $s=+1/2$ (upper curve)  and $s=-1/2$
(lower curve), as a function of time. Fig.\ \ref{Trajectories}(b) shows the
velocities of the defects as the function of their separation $D$.
(See footnote \cite {SYM1} for the simulation parameters.) 

Consider first the dashed trajectories. These were obtained with the
flow field switched off, the case for which equation (\ref{Qevolution}) 
reduces to a Ginzburg-Landau model. As expected the two defects move with 
the same speed and annihilate halfway between their initial positions 
(marked by a horizontal line in Fig.\ \ref{Trajectories}(a)).  The results 
for the Ginzburg-Landau model fit the simple formula (\ref{defectdyneq}) for
$\mu_0=124 \mu$m$^2/$s and $R_c=0.0233 \mu$m for $D\gtrsim R_c$.
Around $D \sim R_c$ the formula overestimates the defect speed.

Consider the effect of the back-flow under the transformation of 
Eq.(\ref{mapping}).  Examining the stress tensor one can see that the 
last term in Eq.(\ref{BEstress}) does not change under the transformation
but the off-diagonal elements of the other terms have their sign inverted.
This reflects the two sources of back-flow in this problem.  
The first has to do with the defect core.
Order is suppressed in the core, which results in an increase
in viscosity at the core (The isotropic viscosity $\alpha_4$, is 
proportional to $(1-q)^2$ where $q$ is the magnitude of the order
\cite{BE94,DO00}).   As a result, the movement of the core
induces similar vortices to those produced when a solid cylinder
is moved through a fluid.  This flow is independent of the
sign of the defect, and points into the direction of 
defect propagation at the core. 
The second source of back-flow comes from
the reorientation of the director field away from the core.
This flow depends on the sign of the spatial derivative of 
the director orientation.  As a result, these two sources of flow
reinforce in one case and partially cancel in the other, thus
giving the anisotropy.  The flow field around the defects is depicted in 
Fig.\ \ref{Twodefects}(b). A strong velocity vortex pair is formed around 
the $+1/2$ defect, with the flow pointing in the direction of defect 
propagation. The flow around the $-1/2$ defect is much weaker, and points
opposite to the direction of defect motion.

The solid line in Fig.\ \ref{Trajectories}(a)
corresponds to a simulation of the full hydrodynamic equations of
motion (\ref{eqBulkFree})--(\ref{BEstress}). There is a marked decrease in the
time to coalescence. This is primarily because the speed of the
$s=+1/2$ defect is increased by $\sim 100\%$ compared to the case
without flow for $D \gtrsim 0.25 \mu$m. The $s=-1/2$ defect is only
slightly affected by the back-flow; the change in velocity is less
than $20\%$.  Due to the speed anisotropy the defects do not meet
halfway between their initial positions.  For defects initially $1 \mu$m apart 
the displacement of the
coalescence point ($\Delta x_1=0.149 \mu$m)  is smaller  than might be
expected from the substantial speed-up of the $s=+1/2$ defect. This is
because the relative flow-induced increase in velocity drops dramatically
near the defect core ($D \lesssim 0.25 \mu$m) where the
defects are moving the fastest.  At these short separations
the relaxational dynamics dominates the hydrodynamics.

Changing the various material parameters of the sample will affect the
velocity of the defects and their speed anisotropy. We find, as
expected, that as the viscosity in the Navier-Stokes equation
increases the motion approaches that of the Ginzburg-Landau
model. Increasing $\Gamma$ in (\ref{Qevolution}) increases the speed
of relaxation to the minimum of the free energy, thus increasing the
speed of the defects.  The speed anisotropy however decreases because
the weight of the free energy relaxation process is increased relative
to the hydrodynamics.  For example the dotted curve in Fig.\
\ref{Trajectories}(a) corresponds to $\Gamma=7.76$ Pa$^{-1}$ s$^{-1}$
(compared to $\Gamma=6.25$ Pa$^{-1}$ s$^{-1}$ for the solid
curve). The displacement of the
coalescence point is $\Delta x_2=0.128 \mu$m $<\Delta x_1$.
Decreasing $A$ in the free energy (\ref{eqBulkFree}) increases the
defect size. The defects move faster but the velocity disparity
decreases, again because the importance of the relaxational dynamics
is increased relative to the hydrodynamics. 

The results in Fig.\ \ref{Trajectories} are for a single elastic
constant (equal Frank elastic constants or, equivalently,
$L_2=L_3=0$). If a more general model for the elasticity is
considered, allowing $L_2 \ne 0$ or $ L_3 \ne 0$, the invariance of
the Ginzburg-Landau equation under the transformation (\ref{mapping})
is broken. Therefore one might expect a difference in the flow
velocities of $s=\pm1/2$ defects even if back-flow is not considered. 

This is indeed the case.  If $L_2=0$ and $L_3<0$ ($L_3>0$) the
$s=+1/2$ ($s=-1/2$)  defect moves faster. For example, if $L_1=8.73$
pN, $L_2=0$, and $L_3=15.88$ pN \cite{SYM2} a comparison of the
velocities $v_{+1/2}$, $v_{-1/2}$ of the $s=1/2$ and $s=-1/2$ defects
respectively gives a speed anisotropy $ a_v = (v_{+1/2}-v_{-1/2})/
(\frac {v_{+1/2}+v_{-1/2}} {2}) \sim -13\%$ at a defect separation
$D=0.5 \mu$m.  For comparison, the anisotropy caused by the back-flow
is $\sim +68\%$. The displacement of the coalescence point is $\Delta
x=0.029 \mu$m. For most materials $L_3>0$ (corresponding to
$K_{11}<K_{33}$), leading to a speed anisotropy opposite to that 
arising from the hydrodynamics.

If $L_2 \ne 0$ and $L_3 = 0$ the velocity anisotropy is very small since for
these values of the elastic constants the relaxational Ginzburg-Landau
dynamics remains invariant under the mapping (\ref{mapping}) in the
limit of uniaxiality and constant magnitude of the order parameter\cite{L3=0}.
The speed anisotropy is small because these conditions are only
relaxed within a defect core.  For example, $L_3=0$ and $L_2 = 15.88$
pN \cite{SYM3} leads to $| a_v | \lesssim 2\% $. 

Anisotropies in the speed of domain walls of up to $50\%$ have been observed
in experiments on pi-cell liquid crystal devices where the movement of twist
and splay-bend walls is important in mediating the formation of the operating
(bend) state from the ground (splay) state. Defects form spontaneously
at these walls and preliminary simulations show that back-flow effects
are responsible for the velocity anisotropy \cite{CT01}. It is also of
interest to investigate the role of defect motion in many other new
generation liquid crystal devices. For example multi-domain nematic
modes improve viewing angles at the expense of introducing defects
into the director profile and understanding  the behavior of such
defects as the electric field is varied will help control device
performance.  In zenithal bistable nematic devices switching is
between two (meta) stable zero-field states. Switching between the
states is mediated by the movement of topological defects \cite{DY01}. 

To conclude, we have used a formulation of nematodynamics based on the tensor
order parameter to study the hydrodynamics of topological defects in nematic
liquid crystals. We find that the coupling between the order
parameter field and the flow has a significant effect on
defect motion: in particular it introduces a substantial difference between
the velocities of defects of different topological charge. Similar but
smaller velocity anisotropies can result from changing the elastic constants.

We would like to thank E. J. Acosta, C. M. Care, J. Dziarmaga, S. Elston, K. Good,
O. Kuksenok, S. Lahiri, N. J. Mottram and T. Sluckin for helpful
discussions. We acknowledge the support of Sharp Laboratories of
Europe at Oxford.  C.D. acknowledges funding from NSF Grant
No. 0083286.

\end{multicols}


\begin{references}
\bibitem{SYM1} Simulations were performed on a $350 \times 350$
  lattice with $A=1.0$, $L_1=0.55$, $L_2=L_3=0$, $\Gamma=0.625$,
  $\xi=0.59$, $\rho=2.0$, $T=0.5$ and $\gamma=3$.
  Periodic boundary conditions were used in
  the $x$ direction and free boundaries were used in the $y$
  direction.
  Given suitable pressure, length and time scales these 
  parameters can
  be mapped to a lattice size $4.4\mu$m$ \times 4.4\mu$m, $L_1=8.73$
  pN, $\Gamma=6.25$ Pa$^{-1}$s$^{-1}$. The Frank elastic constants are
  $K_{11}=K_{22}=K_{33}=2L_1q^2=4.37$pN and Miesowicz viscosities
  $0.02$ Pa s to $0.1$ Pa s. The material parameter values chosen are
  close to those of 5CB.
\bibitem{L3=0} If $L_3=0$ and the director is confined to the $xy$ plane,
the Frank elastic free energy density (assuming uniaxiality and constant
magnitude of order) can be written as $f = \frac {\kappa} {2} (\nabla
\theta)^2$, where $\theta$ is the angle of the director to the
horizontal axis and $\kappa=(2L_1+L_2)q^2$.  The dynamics of the
medium, $\partial_t \theta= - \Gamma \delta F / \delta \theta= \Gamma
\kappa \nabla^2 \theta$, is invariant under mirroring on the $x$
axis. Moreover a given $L_1$ and $L_2$ correspond to the same dynamics
as $L_1'=L_1+L_2/2$ and $L_2'=0$.  (Note that there is no speed anisotropy 
in the latter case with the Ginzburg-Landau model for the tensor
order parameter.)  For defects this equivalence is only approximate
due to the biaxiality and the change in the magnitude of order in the core.
\end{references}
\end{document}